\documentclass[epjCONF]{svjour}
\usepackage{graphics}
\usepackage[varg]{txfonts} % Times fonts
\usepackage[latin1]{inputenc}
\newcommand{\Msun}{\mbox{$M_{\odot}$}}

\session-title{Ageing low-mass stars: from red giants to white dwarfs}
\begin{document}
\title{Red Giant evolution and specific problems}

\author{Alessandro Bressan\inst{1}\fnmsep\thanks{\email{alessandro.bressan@sissa.it}}
\and Paola Marigo\inst{2}
\and L\'eo Girardi\inst{3}
\and Ambra Nanni\inst{1}
\and Stefano Rubele\inst{3}}

\institute{SISSA, via Bonomea 265, I-34136 Trieste, Italy \and
           Dipartimento di Fisica e Astronomia Galileo Galilei,
           Universit\`a di Padova, Vicolo dell'Osservatorio 3, I-35122 Padova, Italy \and
           Osservatorio Astronomico di Padova, Vicolo dell'Osservatorio 5,
           I-35122 Padova, Italy}
\abstract{In spite of the great effort made in the last decades
to improve our understanding of stellar evolution,
significant uncertainties still remain due to our poor knowledge of some
complex physical processes that still require an empirical calibration, such as
the efficiency of convective heat transport and interior mixing.
Here we will review the impact of these uncertainties on the evolution
of red giant stars.}

\maketitle

\section{Introduction}
\label{intro}
Thanks to the efforts of many different groups in the last decades,
stellar evolution has now reached a high degree of
accuracy and completeness.
Indeed, it can now account for a variety of internal physical processes,
follow the most advanced phases and deal with different
chemical compositions, so that one could in principle reproduce
any stellar environment disclosed by the continuously advancing observational
facilities. At the same time observations themselves have become more
and more detailed,  even providing direct access to star interiors, like in the case of
asteroseismology, thus  posing a real challenge to theory.
In spite of these efforts,
physical processes exist that, because of their complexity,
still suffer of large uncertainties with a straightforward drawback:
models lose their predictive power.
This is crucial when dealing with  advanced evolutionary phases,
e.g. the Red Giant Branch (RGB) and the Asymptotic Giant Branch (AGB),
and  stellar populations that are not well represented
in the solar vicinity or in our Galaxy. In fact, we know that about half of the stars
in Universe were formed in elliptical galaxies under very different environmental conditions.
From the theoretical point of view there are key questions which
still lack a definitive answer, namely:
Does the mixing length parameter depend on  metallicity~?
Which is the efficiency of convective overshoot~?
How much does mass loss on the RGB depend on metallicity~?

This review is not intended to tackle these issues,
but rather it will summarise the current theoretical situation with particular emphasis
on evolved stars during the RGB and He-burning phases, for which the above
uncertainties become more critical. We may expect that adding new dimensions to the HR diagram,
such as those provided  by asteroseismology, could allow the biggest improvements just
where the uncertainties are the largest.

\section{The {\rmfamily \small PARSEC} code}
Here we briefly review the basic physics input
used to compute stellar evolution models. We refer in particular to
the new code developed in Padova, PARSEC
({\sl\,P}adova {T\sl\,R}ieste {\sl\,S}tellar {\sl\,E}volution {\sl\,C}ode),
with which we obtained many of the results presented here.
A detailed description can be found in \cite{Bressan12} %  Bressan et al. (2012).
\subsection{Equation of state, nuclear reaction rates, opacities and neutrino emission}
The equation of state (EOS) is computed with the FreeEOS code
developed and updated over the years by A.W.~Irwin \footnote{http://freeeos.sourceforge.net/}.
The FreeEOS package is fully implemented in our code and we may use it
``on-the-fly'' but, since the pre-tabulated version is much faster and
sufficiently accurate for most of our purposes, we proceed by pre-computing suitable tables and
by interpolating between them. As shown by Bressan et al., differences between
the two methods are negligible.

Opacities in the high-temperature regime, $4.2 \le \log(T/{\rm K}) \le 8.7$, are
obtained from the Opacity Project At Livermore
(OPAL) team \cite{Igle}
while, in the low-temperature regime,
$3.2 \le \log(T/{\rm K}) \le 4.1$, we use opacities generated with our
\AE SOPUS\footnote{http://stev.oapd.inaf.it/aesopus} code \cite{Marigo09}.
Conductive opacities are included following \cite{Itoh08}.
The nuclear reaction network consists of the p-p chains, the CNO tri-cycle, the
Ne--Na and Mg--Al chains, and the most important $\alpha$-capture
reactions, including the $\alpha$-n reactions.  The network solves for
the abundances of $26$ chemical species: $^1$H, D,
$^3$He, $^4$He,$^7$Li, $^8$Be, $^4$He, $^{12}$C, $^{13}$C, $^{14}$N,
$^{15}$N, $^{16}$N, $^{17}$N, $^{18}$O, $^{19}$F, $^{20}$Ne,
$^{21}$Ne, $^{22}$Ne, $^{23}$Na, $^{24}$Mg, $^{25}$Mg, $^{26}$Mg,
$^{26}$Al$^m$, $^{26}$Al$^g$, $^{27}$Al, $^{28}$Si.  In total
we consider $42$ reaction rates, that are taken from
the recommended rates in the JINA reaclib database
\cite{Cyb10}
from which we also take the corresponding $Q$-values.
The electron screening factors for all reactions are those from
\cite{Dewitt} and \cite{Grab}.

The abundances of the various elements
are evaluated with the aid of a semi-implicit
extrapolation scheme,  without
assuming nuclear equilibrium and requiring the conservation of the
total number of nucleons during the evolution \cite{Marigo01}.
Energy losses by electron neutrinos are taken from
\cite{Mun} and \cite{Itoh83},
but for plasma
neutrinos, for which we use the fitting formulae provided by
\cite{Haft}.

The energy transport in the convective regions is described according
to the mixing-length theory of \cite{BV}.
The super-adiabatic
gradient is maintained until its difference with respect to the
adiabatic one decreases below $\nabla_{\rm Element}-\nabla_{\rm Adi}~<~10^{-6}$.
The mixing length parameter $\alpha_{\rm MLT}$
is fixed by means of the solar model calibration described below,
and turns out to be $\alpha_{\rm MLT}=1.74$.

Microscopic diffusion is included following the implementation by
\cite{sala}.  The diffusion coefficients are
calculated following \cite{Thoul} and the corresponding system
of second order differential equations is solved together with the
chemistry equation network, at the end of each equilibrium model.
Diffusion is applied to all the elements considered in the code in the
approximations that they are all fully ionized.
\section{The solar model}
\label{sec_sun}
The comparison with the solar model is a necessary step to check the
quality of the input physics, to calibrate the free parameters that
cannot be directly derived from the theory, such as the Mixing
Length parameter,  and to obtain the initial solar abundance of helium and
metals.
\subsection{The solar partition of heavy elements}
\label{distribution}
The full reference distribution of metals in the Sun consists of $90$ chemical
elements\footnote{A few elements (Po, At, Rn, Fr, Ra, Ac, and Pa) are
  assigned negligible abundances.} from Li to U, with abundances taken
from the compilation by \cite{GS98}(GS98). However for a subset
of species we adopt the abundances recently revised by \cite{CA}
(and references therein, CA11).
According to this abundance compilation, the present-day Sun's
metallicity is $Z_{\odot}= 0.01524$.  This value is intermediate between the
most recent estimates, e.g.  $Z_{\odot}= 0.0141$ of \cite{Lodders}
or $Z_{\odot}= 0.0134$ of \cite{Asplund} (AGSS),
and the previous value of $Z_{\odot}= 0.017$ by GS98.
We refer to Grevesse (this volume) for a discussion of the different estimates.
We remind here that the assumption of a different metallicity for the Sun
bears indirectly on the location of the RG stars,
because it affects directly the calibration of the MLT parameter,
as discussed below.
Other partitions of heavy elements, $\{X_i/Z\}$, have been considered,
following the evidence of different chemical evolution paths in our
own Galaxy and in the outer galaxies (e.g. $\alpha$-enhanced mixtures).
\begin{table}[h]
\begin{center}
\caption{Solar calibration assuming the CA11 abundance of metals.}
\begin{tabular}{l l l r l l l}
\hline\hline
&              Solar data                    &        &        &\vline& \multispan{2}{\hfill Model \hfill}   \\
\hline
                              & Value$_\odot$    & error & source &\vline& Tab-EOS & Fly-EOS \\
\hline
$L$ (10$^{33}$erg\,s$^{-1}$) &   3.846     & 0.005       & \cite{Guen} &\vline&    3.848  & 3.841   \\
$R$ (10$^{10}$\,cm)        &   6.9598    & 0.001    &  \cite{Guen} &\vline& 6.9584   & 6.96112 \\
$T_{\rm eff}$ (K)       &   5778   & 8    &   from $L_\odot$ \& $R_\odot$ &\vline&  5779   & 5775  \\
$Z$                        &   0.01524   & 0.0015    & \cite{CA} &\vline&0.01597    & 0.01595   \\
$Y$                        &   0.2485    & 0.0035   &  \cite{Basu04} &\vline& 0.24787    & 0.24762  \\
$(Z/X)$                    &   0.0207    & 0.0015    &  from $Z$ and $Y$ &\vline&0.02169    & 0.02166   \\
$R_{\rm ADI}/R$            &   0.713     & 0.001    &  \cite{Basu97} &\vline&0.7125      & 0.7129 \\
$\rho_{\rm ADI}$                 &   0.1921      & 0.0001    &  \cite{Basu09})&\vline&0.1887      & 0.1881 \\
$C_{\rm S, ADI}/10^7$cm/s        &   2.2356      & 0.0001     & \cite{Basu09} &\vline&2.2359      & 2.2364  \\
\hline
\end{tabular}
\end{center}
\label{table:scalib}
\par{\small For the Tab-EOS model the derived parameters are $Z_i$=0.01774, $Y_{i}$=0.28, $\alpha_{\rm MLT}$=1.74 and Age=4.593 Gyr.
Assuming for the Fly-EOS the same initial parameters, we get an age of 4.622 Gyr. Age includes the pre main sequence phase.}
\end{table}
With the scaled-solar partition of heavy elements $\{X_i/Z_{\odot}\}$ ,
we have computed a large grid of solar models
from the PMS phase to an age of 4.8~Gyr, varying the initial
composition, $Z_{\rm initial}$ and $Y_{\rm initial}$ and the
mixing length parameter $\alpha_{\rm MLT}$.
To calibrate these free parameters, the models
were compared with a set of solar data obtained from the literature
which are summarized in Table~\ref{table:scalib}.  The reference solar
data used here come from MDI observations described in
\cite{Basu00}.  More recent data can be found in
\cite{Basu09} from the BiSON experiment.  However since we intend to
consider the effects of changing the present-day surface solar
composition, we have found it more convenient using the  \cite{Basu00}
data, with respect to which  several comparisons  at different
chemical compositions are available in literature.  Since an important
goal of the comparison is to obtain the mixing length parameter
$\alpha_{\rm MLT}$ that will be
used to compute all other stellar evolutionary sets,
the calibration has been performed exactly with the same set-up used
for the calculations of the other tracks, i.e. with tabulated EOS and
opacities and, of course, using microscopic diffusion.  For sake of comparison
we have also computed a solar model with the ``on-the-fly'' version of FreeEOS,
adopting the parameters of the best fit obtained
with the tabulated EOS, but changing the solar age in order to
match as well as possible the solar data. The parameters of our best
model are also listed in Table~\ref{table:scalib}.

\begin{figure} %\includegraphics{MDI_TLM7_dc2_c2_S11_2012Z0.01774Y0.28ML1.74EO0.05.ps}
\begin{minipage}{0.45\textwidth} %\noindent a)
\resizebox{\hsize}{!}{\includegraphics{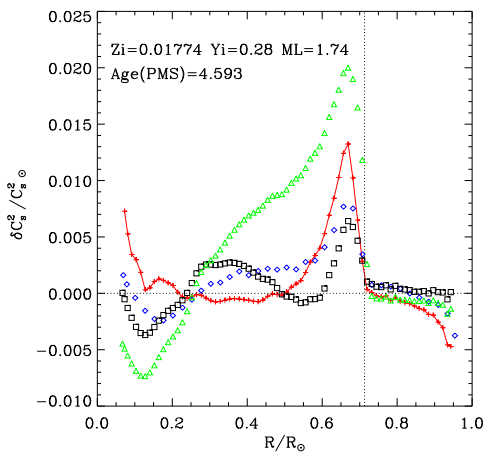} }
\end{minipage}
\begin{minipage}{0.45\textwidth} %\noindent b)
\caption{Solar model obtained with the CA11 abundances
   compared with other calibrations in literature.
   Interior relative differences
   in the squared sound speed $\delta c_{\rm s}^2/c_{\rm s}^2$,
   in the sense Sun minus model.  Solar
  values are obtained from MDI data \cite{Basu00}.  Symbols are
  as follows.  Red crosses: our model; blue diamonds:
   model differences with respect to MDI data by \cite{Basu09}
  using GS98 solar abundances; black squares:
  model by \cite{sere} using GS98 solar
  abundances; green triangles: model by \cite{sere} using
  AGSS solar abundances.
  }
\label{fig_calib}
\end{minipage}
\end{figure}

In Fig.~\ref{fig_calib} we plot the relative
variation of the squared sound speed
$\delta c_{\rm s}^2/c_{\rm s}^2 =
(c_{{\rm s},\odot}^2 - c_{{\rm s, model}}^2)/c_{{\rm s},\odot}^2 $,
as a function of the fractional radius inside the Sun (red crosses and
solid line).
The solar values were obtained from MDI data
\cite{Basu00}.  In the same figure we also show: the
\cite{Basu09} model differences with respect to MDI data,
obtained with the GS98 solar abundances (blue
diamonds); the \cite{sere} difference profile using
GS98 (black squares) or AGSS
 solar abundances (green triangles), respectively.
The quoted age includes the PMS lifetime, here defined as the time elapsed
until when the total gravitational luminosity first goes to
zero on the ZAMS. The PMS lifetime amounts to  about 40 Myr.
We see that our solar model
performs fairly well, also in consideration of the fact that lowering the solar
metallicity relative to the GS98 value was generally found to worsen
the comparison  between the models and the solar data.

Particularly encouraging are the small values of the $\delta c_{\rm s}^2/c_{\rm{s}}^2$
in the central radiative regions of the Sun,
indicating that a good agreement is feasible also with abundances
lower than those of GS98.  Towards the central
regions our model predicts sound speed  and density somewhat
lower than the measured solar values. A similar, though
less pronounced problem, is also present in the comparison of the models
with the solar data extracted from the Bison experiment
\cite{Basu09}.
For the track computed with the ``on-the-fly'' FreeEOS we did not
repeat the whole calibration process but
we may also obtain a fairly good agreement
with the solar data, by slightly increasing the age.
From the initial values of helium and metallicity of the
Sun, $Y_{\rm initial}, Z_{\rm initial}$ (in mass fractions),
and adopting for the primordial He abundance $Y_{\rm p }=0.2485$
(\cite{Kom}), we obtain also the helium-to-metals
enrichment ratio, $\Delta Y/\Delta Z=1.78$. This law has
been used by \cite{Bressan12} to compute several evolutionary sets
at different metallicities.
\section{The evolution after the Main Sequence}
%\begin{figure}
%% For example, with the option graphics use
%\resizebox{0.75\columnwidth}{!}{%
%\includegraphics{bre_fig1.eps} }
%  \caption{}
%    \label{fig_masses}
%\end{figure}
%%
Besides the choice of the microscopic physics, the evolution after
the main sequence is affected by mixing processes related to
the efficiency of convective core overshoot, atomic diffusion and rotation.
For the latter, not yet implemented in PARSEC, we refer to
the discussion  by Eggenberger in this volume.
In PARSEC we adopt a maximum overshooting efficiency  $\Lambda_{\rm~max}=0.5$,
i.e.\ a moderate amount of overshooting, which coincides
with the values adopted in the previous \cite{Bertelli} and
\cite{Gir00} models.  This corresponds to about $0.25\,H_P$
of overshoot region {\em above} the convective border found in other
common formalisms.
It is well known that there are theoretical difficulties in
defining the efficiency of overshooting in the transition region
between stars with radiative cores and those with convective cores
(e.g. \cite{Aparicio}).  In this region,
say between  $M_{\rm\,O1}\leq~M\leq~M_{\rm\,O2}$ where $M_{\rm\,O1}$ and $M_{\rm\,O2}$
depend from the chemical composition,
we assume that the overshooting efficiency $\Lambda_{\rm~c}$
increases linearly with mass from zero to the maximum value.
$M_{\rm\,O1}$ is defined as the initial stellar mass that
maintains a persistent convective core during H-burning.
Practically, $M_{\rm\,O1}$ is the minimum mass of a star in which a convective core
is still present even after its central hydrogen has decreased by a significant amount
($X_{\rm c}\sim$0.2)
from the beginning of the main sequence.
$M_{\rm\,O2}$ is set equal to $M_{\rm\,O1}$+0.3$M_\odot$.
This choice is supported by
the modelling of the open cluster M\,67 (see
\cite{Bressan12}), which indicates an
overshooting efficiency $\Lambda_{\rm c}\simeq0.5$ already at
masses of $\sim\!1.3$~\Msun\ for solar-metallicity stars;
and by the SMC cluster NGC~419 (\cite{Girardi09}, \cite{Kam}),
in which the turn-off probes masses between $\sim\!1.65$ and
1.9~\Msun.
Recent indications come also from asteroseismology but they are still
ambiguous: while the observations of $\alpha$\,Cen~A
(\cite{Meu}) suggest negligible overshooting in
solar-metallicity stars of mass $\sim\!1.1$~\Msun, recent
asteroseismic studies of the nearby old low-mass star HD~203608
(\cite{Deh}), with $[Z/X]\simeq-0.5$,  support the existence
of overshooting (with $\alpha_{\rm ov}=0.17$, which corresponds to
$\alpha_{\rm ov}\simeq0.32$ in our formalism) at masses as low as
$0.95$~\Msun, which is probably just slightly above the $M_{\rm\,O1}$
limit. Clearly, the behavior of overshooting in the transition region
from $M_{\rm\,O1}$ to $M_{\rm\,O2}$ deserves more detailed
investigations.
\begin{figure} %CP_TD_M1.10_Z0.014_Y0.274.ps  fig2a
  \resizebox{\hsize}{!}{\includegraphics{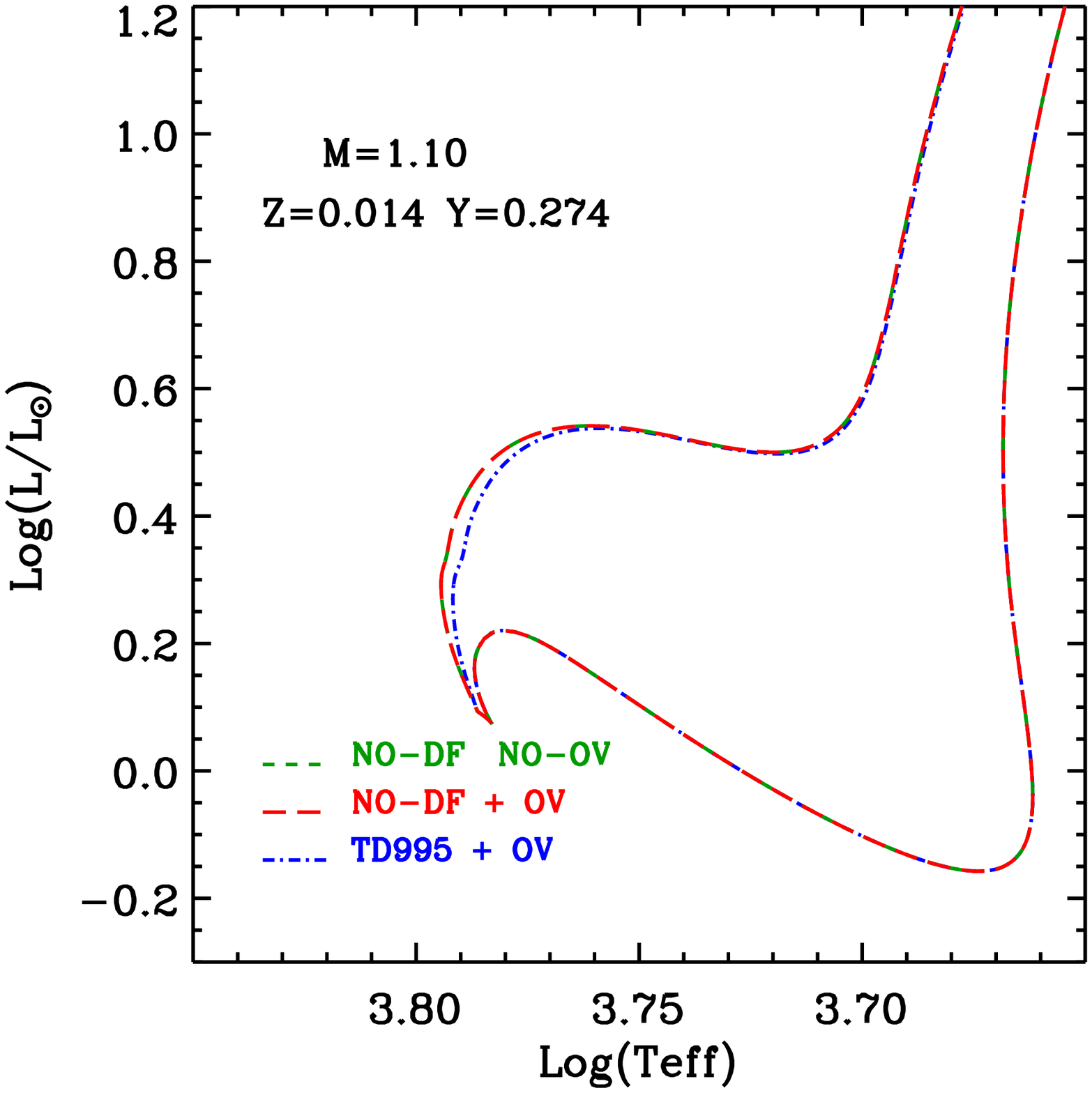}
  \includegraphics{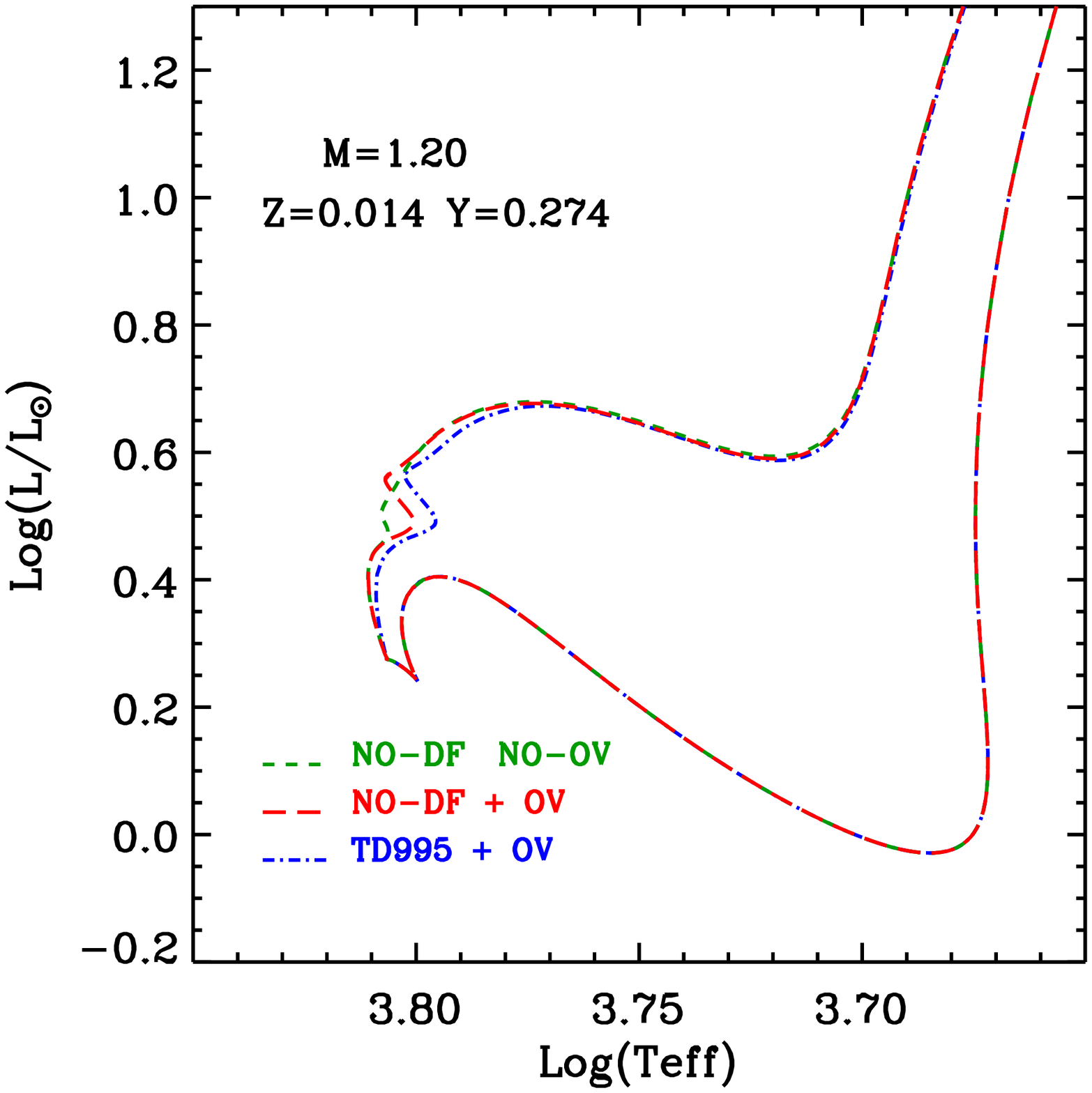}\includegraphics{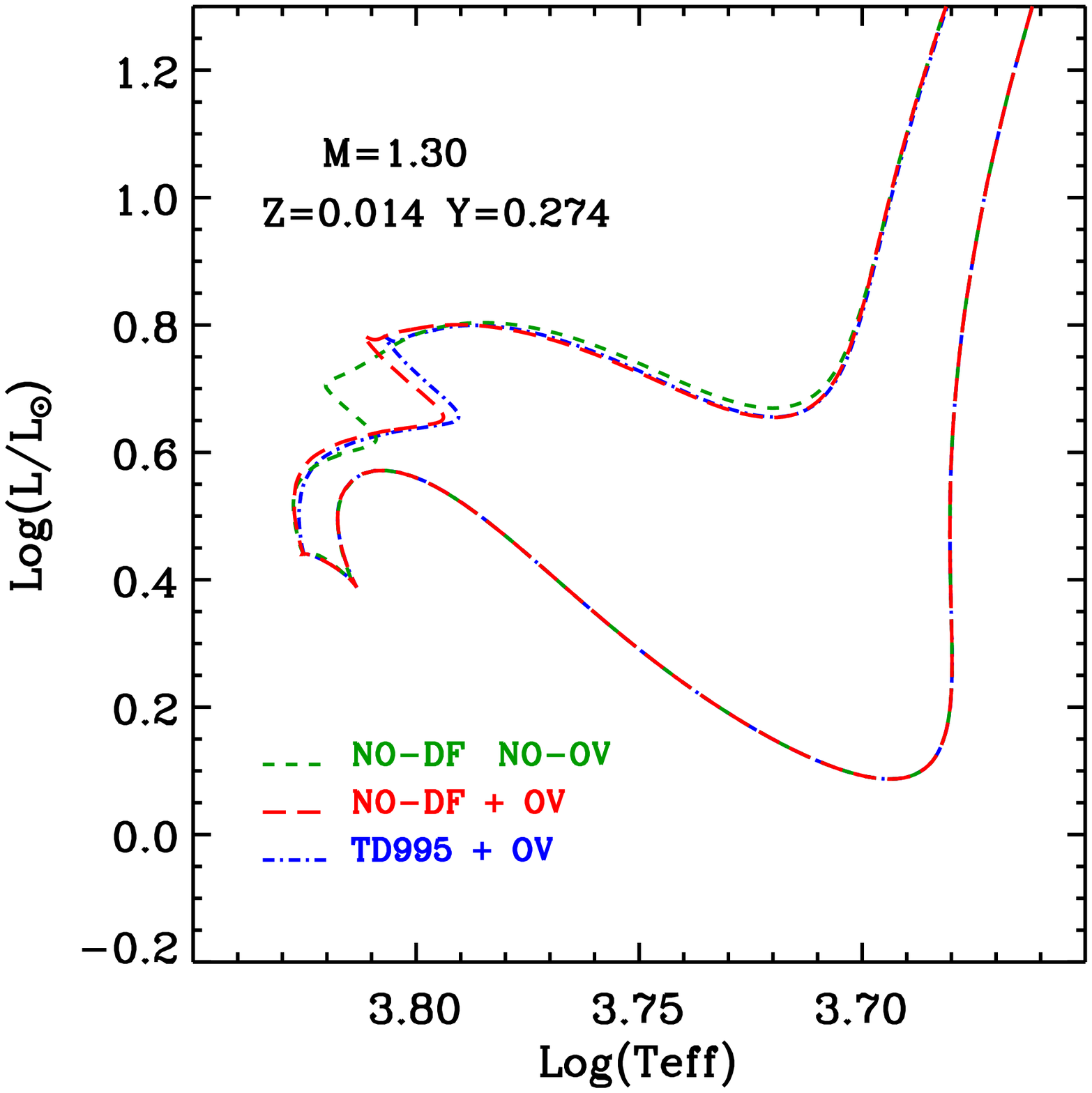}}
\caption{Effects of a different efficiency for
overshoot and diffusion, in the radiative/convective mass transition region. DF stems for diffusion, OV for overshoot
and TD995 for turbulent diffusion in the layers with M$_r$/M $\geq$ 0.995.}
  \label{fig_trans}
\end{figure}
% \subsection{Diffusion and core Overshoot}

Effects of diffusion have also been thoroughly discussed in literature
(see e.g. \cite{Vandb} and references therein).
A still open problem is how to deal with diffusion in stars with
almost fully radiative envelopes. Indeed, on the basis of accurate
abundance determinations in stars of globular cluster NGC~6397
by \cite{Grat}, \cite{Chab} concluded that
microscopic diffusion should be fully inhibited in the external layers
of metal poor stars at least down to a depth of about 0.005~\Msun\
from the photosphere,  and partially inhibited in the region
0.005 to 0.01~\Msun\ from the photosphere. At larger metallicities,
microscopic diffusion in the external
layers of stars of similar mass is already inhibited by the more
extended external convection.  One could think that in
 more massive stars with less extended surface convective
regions, the effects of diffusion are negligible because of the much
shorter stellar evolutionary times.  However
if the mesh spacing in the external layers is kept
suitably small for accuracy purposes, diffusion can noticeably change
the surface composition even for masses well above those of globular
clusters, when external convection disappears (\cite{Turc}).
To cope with these difficulties  some sort of extra-mixing beyond the base of the
external convective layers has been invoked (e.g. \cite{Richer}).
This extra-mixing, of unknown origin but whose effect is that of moderating/inhibiting
other diffusive processes, is parameterized as a turbulent diffusion
with a coefficient that is calibrated on the observed surface
abundances of old stars (\cite{Vandb}).
Unfortunately, when the calibrating observable is the surface Li abundance, the results of different
investigations, based on different stars, do not agree
(\cite{Mel}, \cite{Nor}).  Furthermore, this
calibration is challenged even more by the discovery that early main
sequence stars still suffer from a non negligible mass accretion
(\cite{Dem11}). This tail of accretion, preceded by a very
efficient turbulent mixing during the pre-main sequence phase that
completely destroys Li, could reshape our view of the surface evolution
of this element (\cite{Mol}), and likely requires another
different calibration of the inhibiting mechanism.  For all
these reasons and since we neglect radiative levitation (\cite{Vauc}),
in the present models we have assumed that
{\sl the outermost regions of the star, with M$_r$/M $\geq$ 0.995, are always homogenized}.
In the current version of PARSEC diffusion operates during the main sequence phase throughout the whole star even in presence of a well developed
convective core and it is maintained until the
effects on the evolution of the star become negligible ($M\leq1.6$~\Msun).
This criterion supersedes the one described in \cite{Bressan12} according to which
diffusion was not considered in stars that develop a
persistent convective core, i.e.\ when core overshoot is taken into
account.
Figure \ref{fig_trans} illustrates how the effects of
overshoot and diffusion affect the evolutionary tracks in the radiative/convective
core transition region.
\section{The Red Giant Branch}
\begin{figure}% M1.00BASTI_Z0.0198_Y0.2734.ps fig3
\centering
\begin{minipage}{0.4\textwidth} %\noindent a)
\resizebox{\hsize}{!}{\includegraphics{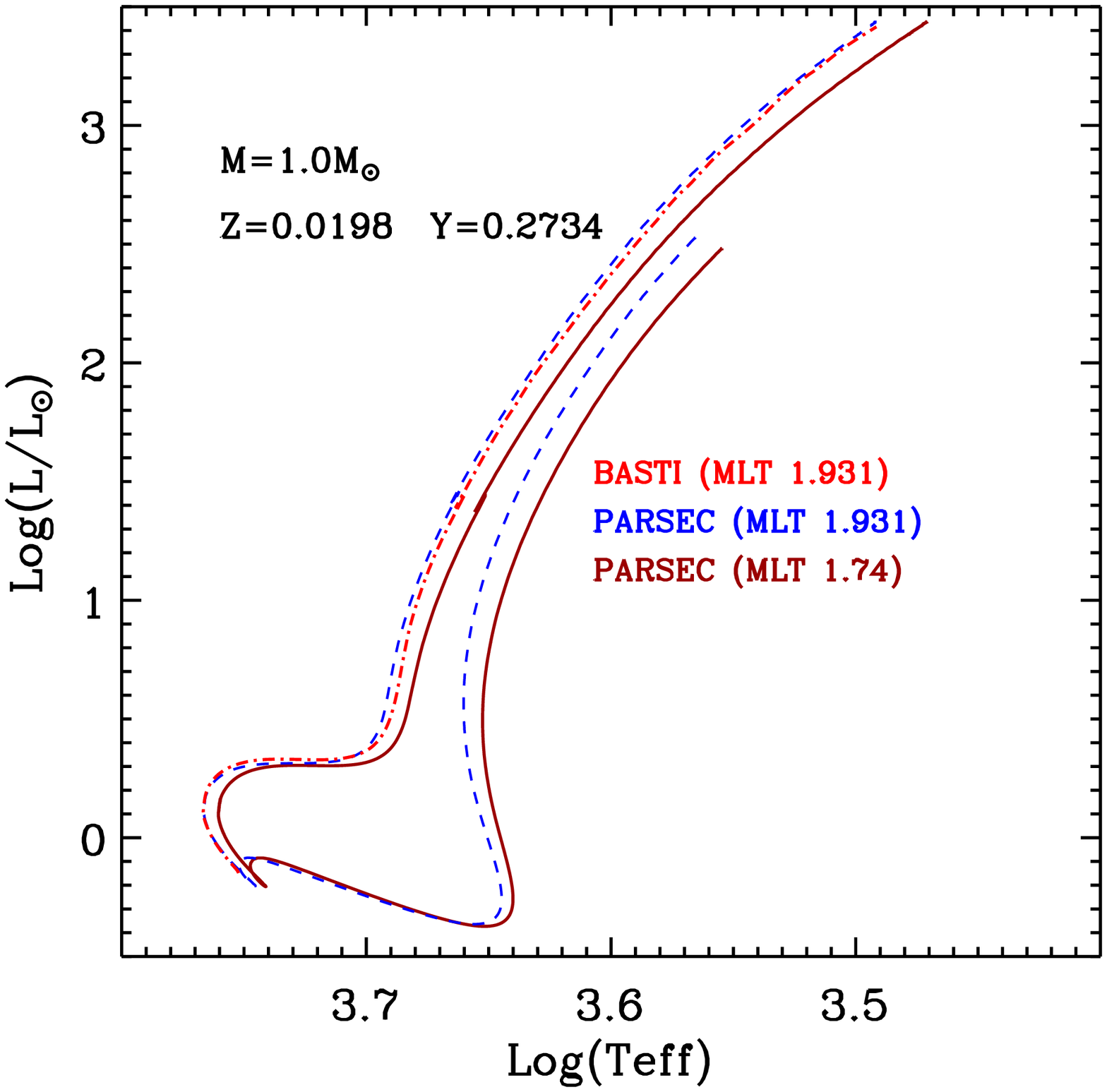}}
\end{minipage}
\begin{minipage}{0.4\textwidth} %\noindent b)
\resizebox{\hsize}{!}{\includegraphics{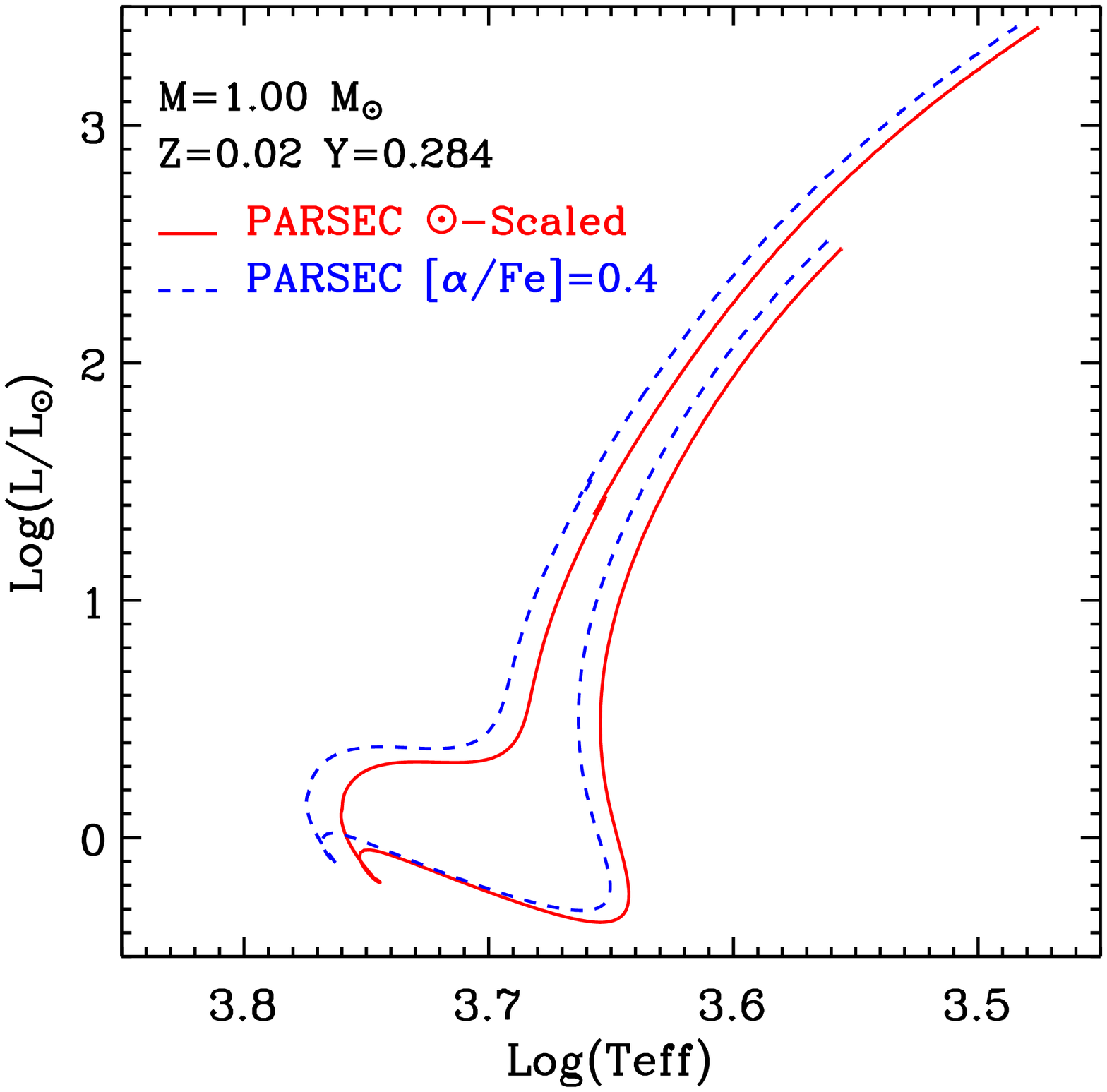}} %M1.00_Z0.02_Y0.284comp_enh.ps}}
\end{minipage}
\caption{Left panel: effects of different solar metallicity
  on the MLT calibration. Right panel: comparison between evolutionary tracks computed either with
  scaled-solar chemical composition, or $\alpha$-enhanced mixtures,
  while keeping the same initial total metallicity Z and helium
  content $Y$.}
\label{fig_enhtracks}
\end{figure}

One of the main goals of stellar evolution is predicting the correct location
(temperature and slope) of the Red Giant Branch (RGB), since it
is used for the derivation of the age of resolved stellar populations, and
provides also a photometric estimate of their metallicity. The RGB is also expected
to importantly contribute to  the integrated colours of old stellar populations.
For a given chemical composition the temperature of the RGB is affected mainly
by  the treatment of inefficient convection (MLT or other models),
the atmospheric boundary conditions, the low-temperature opacities and the EOS.
These effects have been extensively discussed in the literature (e.g. \cite{Mont} % alban et al. 2001,
\cite{Vandb}, \cite{CAS11}). Here we call attention
just to a couple of subtle but important effects.
The first one is that the observed metal abundance of the Sun
impacts directly on the location of the RGB because
of the calibration of the MLT parameter.
The left panel of Figure~\ref{fig_enhtracks} shows that
when adopting the same MLT parameter used by BaSTI, the PARSEC track
of M=1M$_\odot$ runs almost superimposed to the BasTI one for the same composition.
However, the different calibration obtained with the CA11 solar
metallicity results in a lower MLT parameter which
significantly affects the location of the track in the HR diagram.

The other effect is related to the {\sl definition} of
"$\alpha$-enhancement" i.e. to the way the chemical
mixture is effectively built.  In general, given an $[\alpha/{\rm~Fe}]$
ratio, one has two options: a) either keeping the ratio [Fe/H]
fixed while increasing the absolute abundances of the $\alpha$
elements, {\sl which leads to a net increase of the total metallicity $Z$};
or b) keeping the metallicity $Z$ fixed while depressing the
abundances of the Fe-group elements and somewhat enhancing those of
the $\alpha$ elements.   In the
former case the most relevant consequence is the effective increase of
O, Mg, Ne, etc., while in the latter case the most important effect is
the depletion of the Fe-group elements.

The right panel of Figure~\ref{fig_enhtracks} exemplifies the effect of adopting
$\alpha$-enhanced mixtures on the evolutionary tracks in the H-R
diagram, {\sl using the b) option}.  There is a systematic trend of
the $\alpha$-enhanced tracks (blue line) to be somewhat warmer than the corresponding scaled-solar
cases (red line), especially along the RGB.
The changes in the abundances may produce important effects on
evolutionary tracks, mainly due to opacity effects (\cite{Marigo09}).
As a general rule we may expect that $\alpha$-enhanced tracks computed
according to the a) option tend to be cooler than the corresponding
scaled-solar tracks (e.g. \cite{Vandb}) because of a
{\sl net increase in the metallicity}, while $\alpha$-enhanced tracks
computed according to the b) option tend to be cooler than the
corresponding scaled-solar tracks because the depression of the
Fe-group elements implies a reduction of the H$^{-}$ opacity,
important absorption source at temperatures $\approx 3000-6000$ K. In
fact the H$^{-}$ opacity is highly sensitive to the number of free
electrons, a significant fraction of which is just provided by
Fe-group elements.
\begin{figure}
  \centering
  \resizebox{0.8\hsize}{!}{\includegraphics{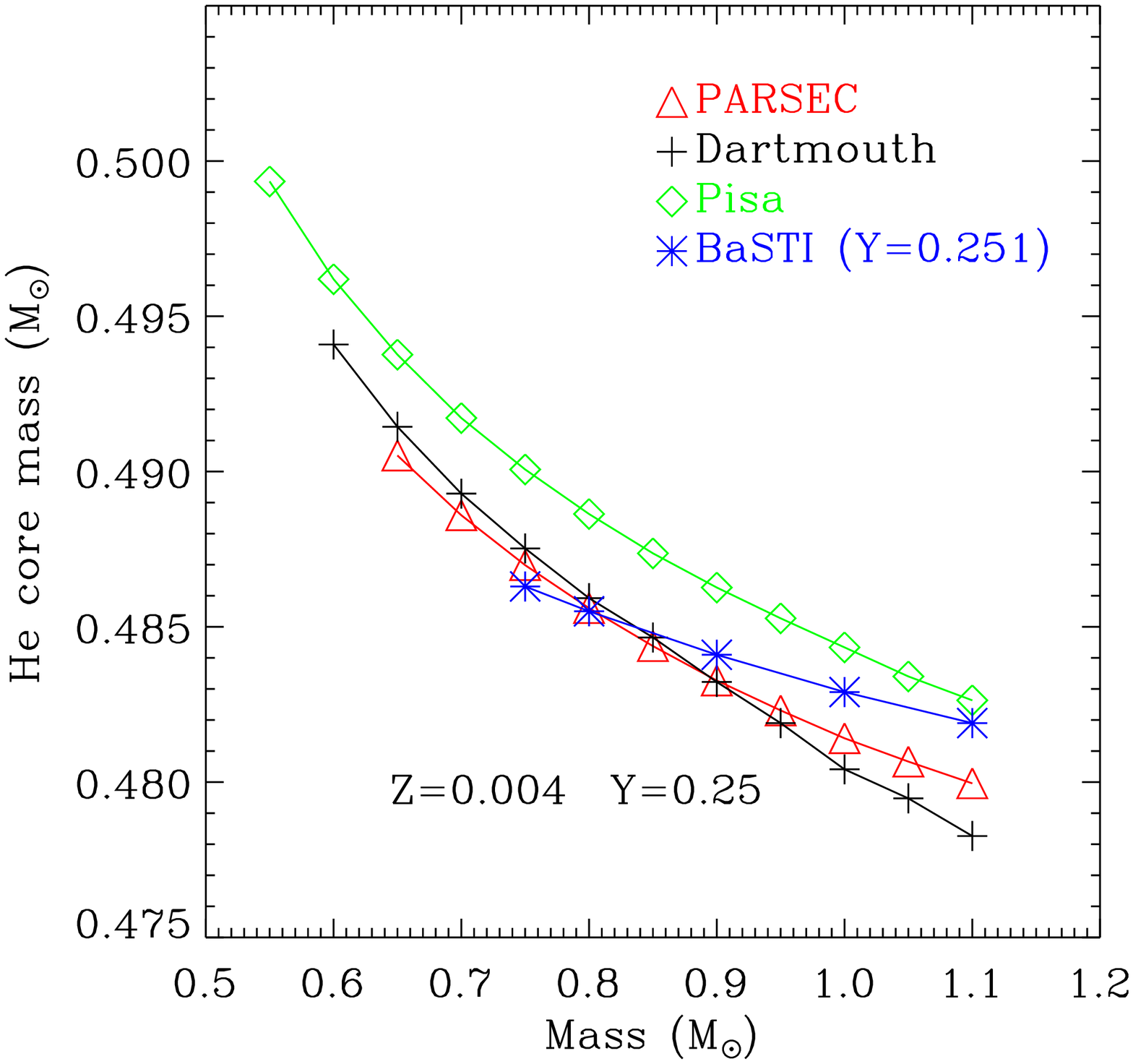} %NV_ND_S12MCORE_Z0.004_Y0.25.PS}
  \includegraphics{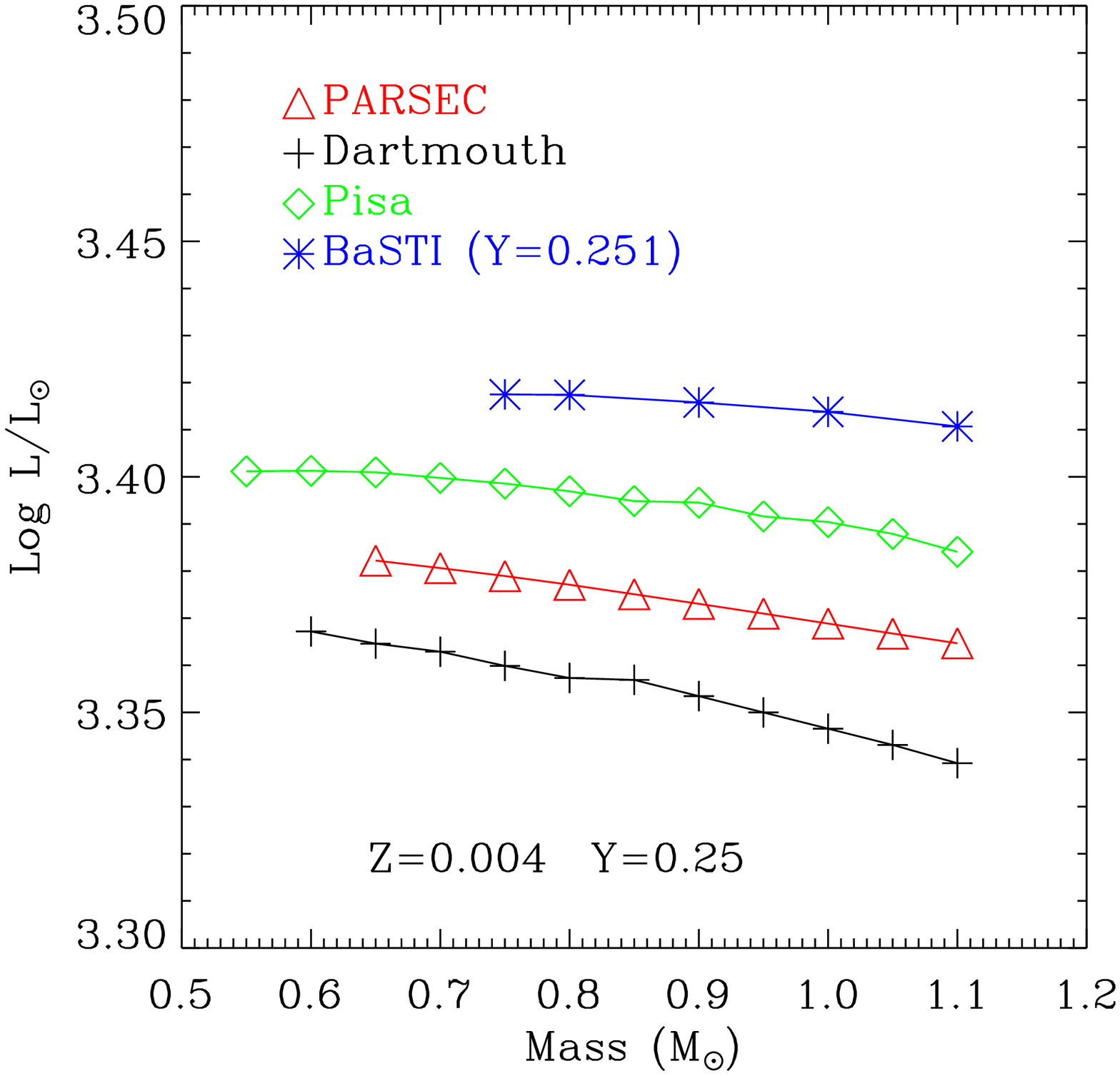}}  %NV_ND_S12LCORE_Z0.004_Y0.25.PS}}
\caption{He core mass (left panel) and luminosity (right panel) at the
  tip of the red giant branch, as predicted from different models,for $Z=0.004,
  Y=0.25$.   Models are from PARSEC,
  Dartmouth, Pisa, and BaSTI.}
  \label{fig_comp}
\end{figure}
\subsection{The RGB Bump}
\label{sec_conv}
During the first red giant ascent low-mass stars
experience a brief though noticeable readjustment of their
internal structure, when the hydrogen burning shell reaches the
chemical discontinuity left by the maximum depth of the
convective envelope during by the first dredge up episode.
This brief stall produce an observable excess of stars
in the CM diagram, named the RGB bump, that can be easily compared
with theoretical predictions. Indeed the observed location of the
RGB bump in globular clusters is about 0.2 to 0.4~mag fainter than
that predicted  by models
(\cite{Dicec10}; see also \cite{CAS11}), though
at the higher metallicities this result depends on the adopted
metallicity scale.
\cite{Alongi} were the first to consider the possibility that
a more efficient mixing at the base of the convective envelope
(envelope overshoot) could easily remedy the above discrepancy
by deepening the location of the chemical discontinuity.
Assuming  an extra mixing of about
$\Lambda_{\rm{e}}=0.5\,H_P$ below the bottom of the convective region,
the RGB bump happens $\sim\!0.3$~mag fainter, which almost fills the
observed discrepancy
It is worth noticing here that another effect of such an extra mixing is
the wider extension of the blue loops of intermediate-mass stars (\cite{Alongi}).
Against the plausibility of such mixing it has always been argued that the
calibration of the solar model does not require a sizable overshoot
region, because the transition between the fully adiabatic envelope and
the radiative underlying region in our Sun is already well reproduced
by models without overshoot. However, this does not exclude the
possibility that just below the fully adiabatic region convection may
penetrate in form of radiative fingers that are able to induce a
significant mixing.  Very recently it has been suggested that a mechanism
of this kind could provide an even better agreement with the physical
state of matter in this transition region derived from solar
oscillations data (\cite{Chr11}).  The size of
this region has been recently estimated to be
$\Lambda_{\rm~e}\sim0.4\,H_P$,  but consistent results may be  obtained
also with a larger value,
$\Lambda_{\rm~e}\sim0.6\,H_P$, which is in very good agreement with the
one adopted since \cite{Alongi}.
These simple considerations would favor larger values of
$\Lambda_{\rm e}$ in low-mass stars.
The fact that the discrepancy disappears at higher metallicities
is intriguing and deserves  a more careful investigation.
In this respect,  the effects of the envelope overshoot
and mass accretion on the abundance of
light elements (e.g. $^7$Li) during and after the PMS evolution
(\cite{Mol}) may provide an critical test against observations.
\begin{figure}[b]
\begin{minipage}{0.45\textwidth} %\noindent a)
\resizebox{\hsize}{!}{\includegraphics{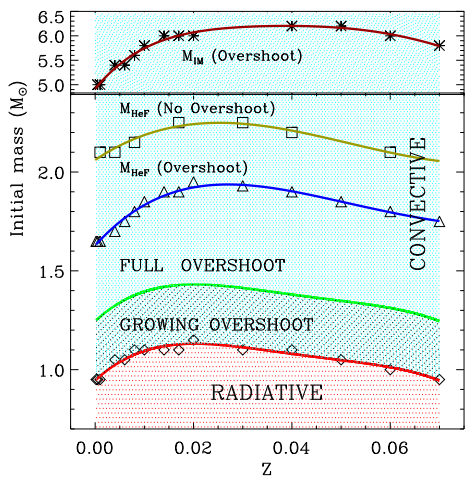} }
\end{minipage}
\begin{minipage}{0.50\textwidth} %\noindent b)
\caption{The behaviour of a few critical masses as a function of
    metallicity, for scaled-solar models following the
    $Y=0.2485+1.78\,Z$ enrichment law.
    From bottom to top: the minimum mass that maintains a persistent
    convective core during H-burning, $M_{\rm\,O1}$;
    the mass above which core overshoot is taken at the
    maximum efficiency $M_{\rm\,O2}$; the minimum mass for a model to
    ignite central He non degenerately, $M_{\rm HeF}$ for both the
    overshoot and the no-overshoot cases; finally, in the upper panel,
    the minimum mass of the stars that ignites C in a non electron
    degenerate core, $M_{\rm\,IM}$. The resolution in
    determining these mass limits is of 0.05~\Msun\ for $M_{\rm\,O1}$
    and $M_{\rm\,HeF}$, and of 0.2~\Msun\ for $M_{\rm\,IM}$. The curves
    are polynomial fits to the corresponding values.
  }
\label{fig_masses}
\end{minipage}
\end{figure}
\subsection{The Tip of the Red Giant Branch}
\label{sec_compbasti}
The  RGB evolution of single low-mass stars ends with the helium flash  inside
electron-degenerate core, if the initial mass is M$\gtrsim$0.5M$_\odot$.
The helium core mass at the Tip of the RGB (TRGB) determines
its luminosity and  as well affects the luminosity of the subsequent helium burning phase.
In nearby resolved galaxies the TRGB can be easily detected and it is one of the most
important distance indicators.

For a given metallicity, the He-core mass and the TRGB luminosity predicted by different models
depend on i) the adopted input physics, i.e. opacities, EOS,
nuclear reactions, neutrino rates and efficiency of atomic diffusion, ii)
the adopted solar chemical composition and
calibration, iv)  and computational details of the previous evolutionary phases.
Figure~\ref{fig_comp} compares the He-core mass and the luminosity at
the TRGB, as predicted by different models in the literature:
PARSEC (\cite{Bressan12}), Dartmouth (\cite{Dotter08}), Pisa (\cite{dellom}) and
BaSTI (\cite{Pietr}).
As for the He-core mass, there is a fairly nice agreement
between different groups. The agreement is slightly less good
for the TRGB luminosities. At M$\sim$0.8~M$_\odot$,  models from Dartmouth,  PARSEC
and BaSTI with the same He-core mass, present luminosities that increase
from the former to the latter.
The difference between Dartmouth and  PARSEC amounts to 0.05~mag while the difference between
Dartmouth and  BaSTI is of about 0.15~mag.
A thorough discussion of possible effects produced by e.g.
conductive opacities and nuclear reaction rates (in particular the
$^{14}N(p,\gamma)^{15}O$ reaction) can be found in \cite{CAS11}.
\section{He-burning stars}
The threshold initial mass below which stars eventually undergo the helium flash, M$_{HeF}$,
depends critically on the chemical composition and on the assumed efficiency of core mixing processes
during the H-burning phase (mainly convective overshoot and rotation).
The behavior of $M_{\rm HeF}$ as a function of
metallicity is shown
in Figure \ref{fig_masses}. Without overshoot, $M_{\rm HeF}$ is about 0.3~M$_\odot$ larger
than for models computed with a mild overshoot efficiency.
In Figure \ref{fig_masses} we plot also
the behavior of $M_{\rm\,O1}$ and $M_{\rm\,O2}$
for scaled-solar models that follow  the
$Y=0.2485+1.78\,Z$ enrichment law. These critical masses may change moderately
by varying the He content at a given metallicity.

\begin{figure}
  \resizebox{0.95\hsize}{!}{\includegraphics{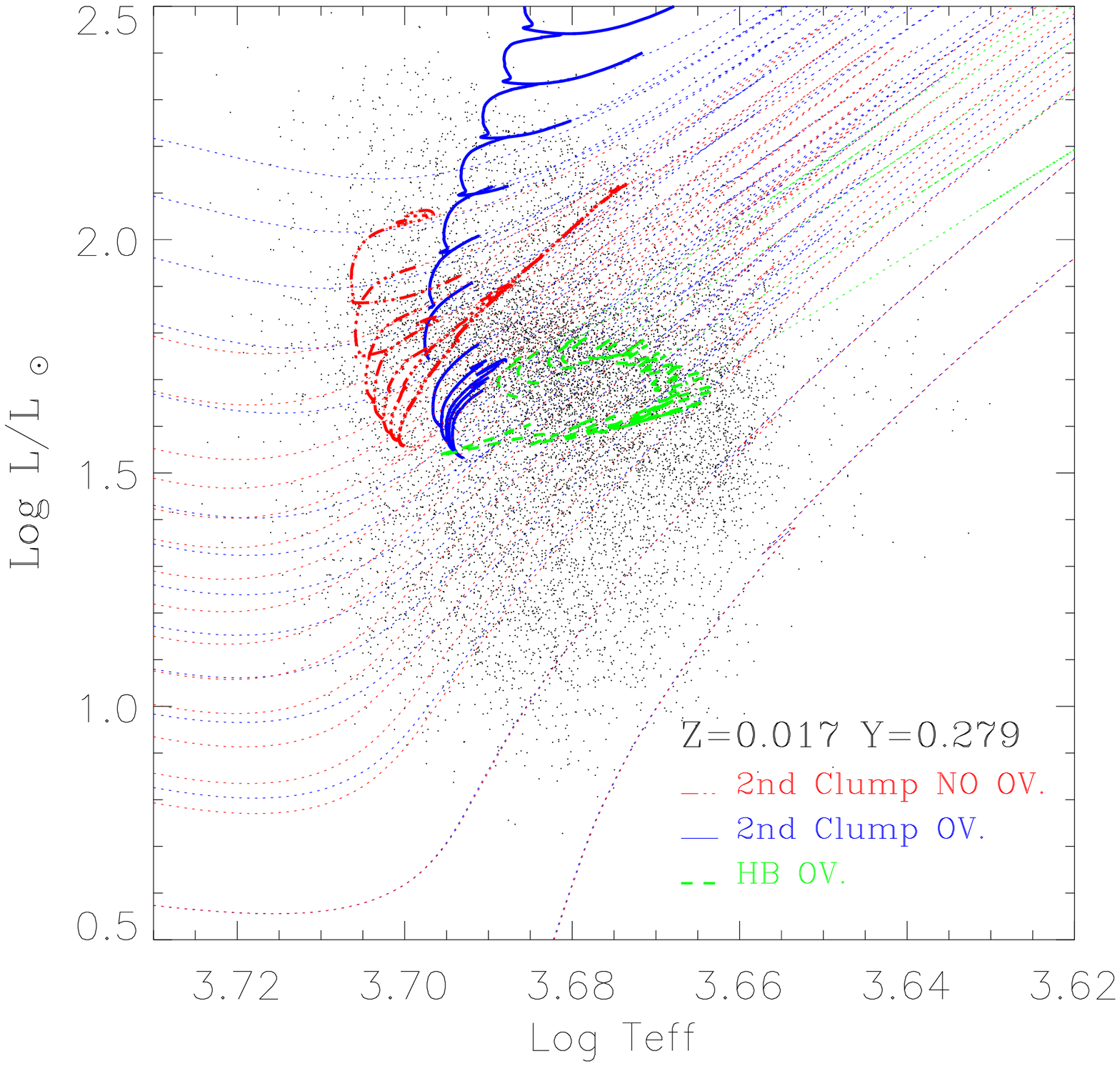}  %OHR.ps}
  \includegraphics{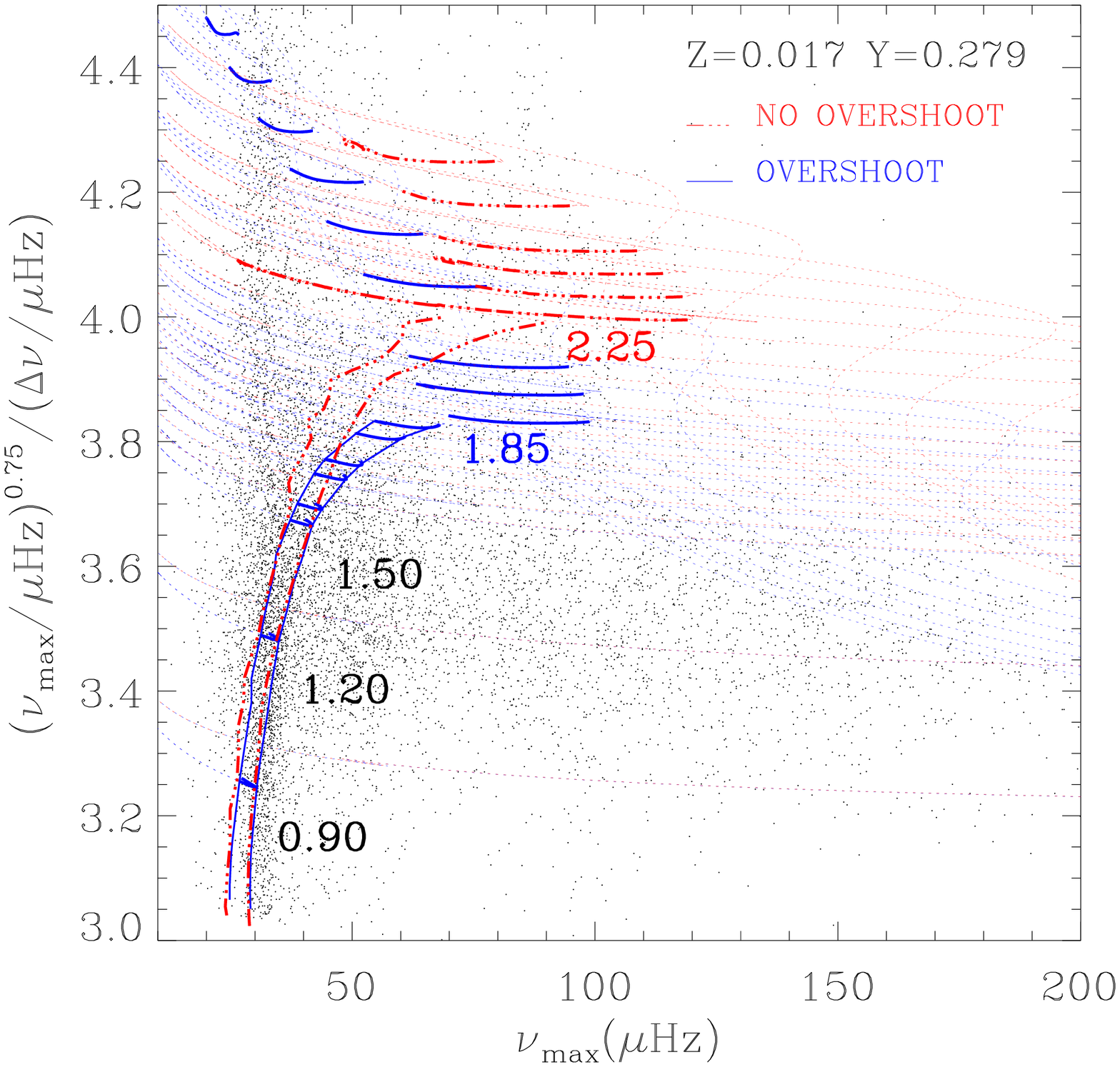}} %Onu.ps}}
\caption{Evolutionary tracks superimposed to the
the Red-Giant stars of the public Kepler data set (\cite{Hek}). The left panel
shows the HR diagram, while
the right panel shows the corresponding frequency domain using, for the models,
the scaling relations by \cite{Hub}.
Green and red dots mark the location of the helium burning stars after the helium flash
for models with ($M_{\rm HeF}\sim$1.85~M$_\odot$) and without ($M_{\rm HeF}\sim$2.25~M$_\odot$)
core overshoot, respectively. The 2$^{nd}$RC tracks are also shown for models with (blu) and without (red) core overshoot.}
  \label{fig_HB}
\end{figure}

The position of central helium burning stars in the HR diagram
is mainly determined by the size of the He-core and H-rich
envelope and of course by the chemical composition.
Stars undergoing the helium flash have similar He-core masses and,
when the envelope is sufficiently small, they occupy the Horizontal Branch (HB)
while, as the envelope mass become sizable, they distribute on a Red Clump (RC) at cooler temperatures.
For the (initial) solar metallicity the HB and RC loci are shown by the
thick green tracks in the left panel of  Figure \ref{fig_HB}.
At lower metallicity the locus extends at much hotter temperatures
for lower initial masses, while it is similar for the higher masses.
If the initial mass is larger than $M_{\rm HeF}$ the stars ignite helium quiescently. Moreover
for M$_i\sim$$M_{\rm HeF}$ they posses {\it a smaller He-core}.
This is why they constitute a separate clump with luminosity
even lower than that of the stars slightly less massive, as shown by the thick blue tracks
in the left panel of  Figure \ref{fig_HB}.
This clump is commonly named the secondary Red Clump  (2$^{nd}$RC;
\cite{Gir99}).
Green and blue tracks in the left panel of  Figure \ref{fig_HB}
are computed taking into account core overshoot.
Tracks without core overshoot are shown in red and only for initial masses
larger than M$_{HeF}$ in order to highlight the location of the 2$^{nd}$RC
for models without overshoot.
The points in Figure \ref{fig_HB} are the Red-Giant stars of the public
Kepler data set (\cite{Hek}).
The models shown in Figure  \ref{fig_HB} do not account for mass loss during the RGB,
which is another source of uncertainty. Indeed recent asteroseismic  determination
of the masses of stars on the RGB and in the RC of the
old metal-rich cluster NGC 6791, indicate that mass loss could be
less efficient than thought before (\cite{Miglio}, \cite{reim}).

The right panel of Figure \ref{fig_HB} shows the models in the frequency domain
after adopting suitable scaling relations and a functional form that
 almost eliminates the dependence on the stellar luminosity
(\cite{Hub}).  In the right panel, green and red dots mark the location of
the helium burning stars after the helium flash
for model with ($M_{\rm HeF}\sim$1.85~M$_\odot$) and without ($M_{\rm HeF}\sim$2.25~M$_\odot$)
core overshoot, respectively.
The corresponding 2$^{nd}$RC tracks of stars with M$>$$M_{\rm HeF}$
are also shown in blue and red, respectively.
\section{Conclusions}
We discussed a number of uncertainties that still affect stellar evolution models.
Our solar model obtained with the CA11 revised solar metallicity
compares fairly well with that computed with GS98 abundances. We have stressed that
a correct estimate of the solar metallicity has an impact also on the advanced evolutionary phases.
While some discrepancies existing in the past have been removed (e.g. at the RGB Tip),
the problem remains of understanding the efficiency of interior mixing
during the main sequence, as well in the later evolutionary phases.
The latter has a significant impact on the turn-off region and on the clump of He-burning stars.
In the latter phase a mild overshoot produces 2$^{nd}$RCs with masses $\sim$0.3~M$_\odot$
lower than without overshoot. Since $M_{\rm HeF}$ depends also on the chemical composition,
Asteroseismic measuremets of RC masses in star clusters of known abundances
will be of paramount importance in resolving this issue.

\bigskip
{\bf Aknowledments} We warmly thanks the organizers for the kind hospitality.

\end{document}